# Deep subwavelength resonant metaphotonics enabled by high-index topological insulator Bi$_2$Te$_3$

*Danveer Singh$^{1,2}$, Sukanta Nandi$^{1,2}$, Yafit Fleger$^{2}$, Shany Z. Cohen$^{1,2}$ and Tomer Lewi$^{1,2*}$*

$^1$Faculty of Engineering, Bar-Ilan University, Ramat-Gan 5290002, Israel
$^2$Institute for Nanotechnology and Advanced Materials, Bar-Ilan University Ramat-Gan 5290002, Israel
$^*$Corresponding authors: tomer.lewi@biu.ac.il

## Abstract

In nanophotonics, small mode volumes, high-quality factor (Q) resonances, and large field enhancements without metals, fundamentally scale with the refractive index and are key for many implementations involving light-matter interactions. Topological insulators (TI) are a class of insulating materials that host topologically protected surface states, some of which exhibit extraordinary high permittivity values. Here, we study the optical properties of TI bismuth telluride (Bi$_2$Te$_3$) single crystals. We find that both the bulk and surface states contribute to the extremely large optical constants, with the real part of the refractive index peaking at n~11. Utilizing these ultra-high index values, we demonstrate that Bi$_2$Te$_3$ metasurfaces are capable of squeezing light in deep subwavelength structures, with the fundamental magnetic dipole (MD) resonance confined in unit cell size smaller than λ/10. We further show that dense ultrathin metasurface arrays can simultaneously provide large magnetic and electric field enhancements arising from the surface metallic states and the high index of the bulk. These findings demonstrate the potential of chalcogenide TI materials as a platform leveraging the unique combination of ultra-high-index dielectric response with surface metallic states for metamaterial design and nanophotonic applications in sensing, non-linear generation, and quantum information.

## Introduction

Metasurfaces have emerged as enabling technology for the development of novel nanophotonic devices and enhanced light-matter interaction platforms, thanks to their unprecedented ability to control the various degrees of freedom of light at subwavelength scales[1–3]. Extensively studied platforms include metallic[4,5], all-dielectric[6], and hybrid[7] (metal-dielectric) metasurfaces and have demonstrated unparalleled wavefront control across the visible to terahertz spectral range. Metallic metasurfaces supporting surface plasmon excitations, inevitably exhibit substantial ohmic losses and are mostly preferred for applications requiring large field enhancements and field localization[8]. In contrast, all-dielectric metasurfaces are the dominating platform for lossless high-efficiency meta-optic devices[9–11]. All-dielectric structures inherently support both electric and magnetic multipole excitations[11], enable a convenient platform for active tunability through various physical phenomena[12–18], and can easily be integrated with existing technologies and CMOS fabrication processes[1]. A rich library of optical functionalities have been demonstrated in various platforms such as spectral and spatial filtering [6,7,19–22], polarization control [23,24], beam focusing[25–29], beam deflectors[30–32], holograms[33,34], image processing[35–37], and nonlinear phenomena[38–42], to name a few. In most cases, extensively studied semiconductor materials are being used, having moderately high refractive indices such as Si ($n \approx 3.7$), GaAs (3.5) and Ge ($n \approx 4$), whereas higher optical constants can be found in more exotic materials systems such as $WS_2$ and $WSe_2$ ($n \approx 5.3$)[43–45], PbTe ($n \approx 6$), or GeSbTe[13,46,47] ($n \approx 6.4$). The ability to localize light in subwavelength unit cells, with large field enhancements and narrow linewidth, fundamentally scale with the optical constants. Thus, the refractive index puts an upper limit to the size of the meta-atom unit cell and hence the overall metasurface spatial resolution. To develop more compact and further advanced nanophotonic devices, material systems with high refractive indices are required.

Topological insulators(TIs) have the potential to provide strong dielectric response with extreme polarizability and have been shown to possess very high refractive indices[48–51]. TIs are materials which are insulating in the bulk while their surface always hosts high mobility conducting states[52]. Moreover, these metallic surface states are time-reversal symmetry protected[53,54], characterized by strong spin-orbit coupling[55] and are robust against defects[56], leading to dissipation-less carrier transport. Chalcogenide TIs are particularly interesting as they possess narrow bulk bandgap[57,58], exhibit extremely high permittivity values, strong anisotropic behavior[59] and are very attractive in the fields of electronic[60], spintronics[61], and photonic applications[62]. Recent demonstrations such as non-integer high harmonic generation[49], plasmonic properties in the ultraviolet to visible range[63,64] along with the high mobility surface states in the mid-infrared (MIR) and THz spectral rages[65–67] indicate the potential of TIs[49,55,60–63,65–67].

Here, we study the optical properties of single crystal TI bismuth telluride ($Bi_2Te_3$). Using infrared spectroscopy along with multiple Lorentz oscillator modeling, we extracted the complex permittivity and refractive index across the NIR to MIR spectral ranges. We obtain record high refractive index value n~11 at MIR wavelengths, substantially higher than other materials. We utilize this ultra-high refractive index to fabricate deep subwavelength nanostructures and demonstrate that the fundamental magnetic dipole (MD) mode can be resonantly excited in metasurface unit cells which are an order of magnitude smaller than the free space wavelength ($<\lambda/10$). We further show that dense ultrathin metasurface arrays provide large electric and magnetic field enhancements arising from the surface metallic states and the ultra-high index of the bulk. These findings highlight the potential of chalcogenide TIs as a diverse platform leveraging the unique combination of ultra-high-index dielectric response with surface plasmonic states and may open up avenues for developing novel ultracompact nanophotonic devices.

## Results

**Optical constants of single crystal $Bi_2Te_3$**

The optical constants of $Bi_2Te_3$ single crystals were derived from infrared spectroscopic measurements of double-sided polished samples. Reflectance spectra were recorded using a fourier transform infrared (FTIR) spectrometer in the 1µm – 18µm spectral range. To analyze the experimental data and extract the dielectric function and the complex refractive index of $Bi_2Te_3$, we used RefFIT program[68] in which a combination of Drude-Lorentz oscillators were used to model the material response. A typical reflectance measurement from a 1 mm thick $Bi_2Te_3$ sample is presented in Figure 1a (red curve). The absolute reflectance (see supporting information for more experimental details) exhibit unusually high reflectivity for non-metallic materials and has a non-trivial spectral dependance. The small dip in the reflectance spectra at 1160cm$^{-1}$ ($\lambda$=8.62µm, marked by the circle in figure 1a) is attributed to the bulk bandgap. It is consistently observed in our measurements and is in-line with previous studies[49,57]. The best fitted reflection curve plotted in Figure 1a (black line) is in very good agreement with the experimentally measured reflectance. This fitted model includes three Lorentz oscillators for the bulk response and one Drude oscillator to account for the topological metallic surface and was used to extract the complex refractive index shown in Figure 1b. All model parameters are summarized in table 1 of the supporting information.

The dispersion of the experimentally extracted complex refractive index plotted in Figure 1b begins with a spectral band with dominant absorption k~n in the short wavelength NIR, followed by a broad MIR range with low-loss dielectric response and ultra-high polarizability. In this MIR range the real part of the refractive index peaks at n≈11 for $\lambda$=6.3µm. This record high index value is in-line with previous studies[48,49,57] and demonstrates the potential of $Bi_2Te_3$ as an ultra-high index dielectric material in the MIR range. Moreover, from Figure 1b we can also identify a broad spectral band of high refractive index and low losses between 6µm to 18µm. This MIR spectral band highlights the region where $Bi_2Te_3$ is mostly attractive for the development of ultracompact nanophotonic devices.

The Drude-like decent in the real part of the index at longer wavelength (Figure 1b) is attributed to the metallic states with possible contribution from bulk carriers from unintentional doping and is further discussed the supporting information.

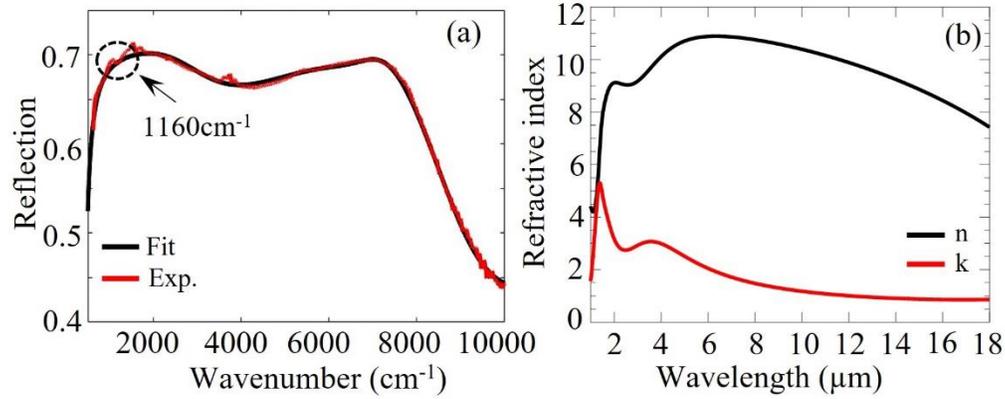

**Figure 1:** Reflectance and optical constants of $Bi_2Te_3$ single crystals (a) Absolute experimental infrared reflectance of $Bi_2Te_3$ single crystal (red). The black line presents the best fit to the experimental curve using multiple Lorentz oscillator model and optimized using RefFIT software. (b) The complex refractive index of $Bi_2Te_3$ across the 1-18µm infrared spectral range. Experimentally extracted complex refractive index obtained from RefFIT showing peak refractive index n≈11 and a broad 6-18µm range of very high refractive index values and low-losses.

To gain a better understanding on how the bulk and the topological surface states contribute to the dielectric function, we also used an analytical two-layer model system[69,70]. In this system the topological surface and the dielectric bulk are modeled by a thin metallic layer (of thickness 1.5nm) sitting atop a bulk dielectric substrate, respectively (see supporting information Figure S2). In terms of contribution to the dielectric function, the bulk part of the TI $Bi_2Te_3$ is modeled by a Lorentz oscillator whereas the topological surface layer is modeled by a single Drude oscillator [71,72]. Such a model system has been successfully used to retrieve the optical constants of topological insulator systems[64].

From this analysis, the contribution of surface states to the Drude like decent in the refractive index, can be clearly seen (Figure S2 of supporting information). With no metallic surface present, the refractive index would have remained constant at peak values (n≈11) for long MIR wavelength range. The strong dielectric response and peak polarizability of $Bi_2Te_3$ in the MIR range culminating in record-high refractive index values, are very intriguing for ultracompact nanophotonic devices with extreme manipulation capabilities and are the basis for the following sections demonstrating deep subwavelength $Bi_2Te_3$ metasurfaces.

**Resonant modes of TI $Bi_2Te_3$ metasurfaces**

The ultra-high permittivity and refractive index values of $Bi_2Te_3$ in the MIR are extremely valuable as they enable tight light confinement[58] with small mode volumes, and large field enhancements, and are key for many implementations involving light-matter interactions[73].

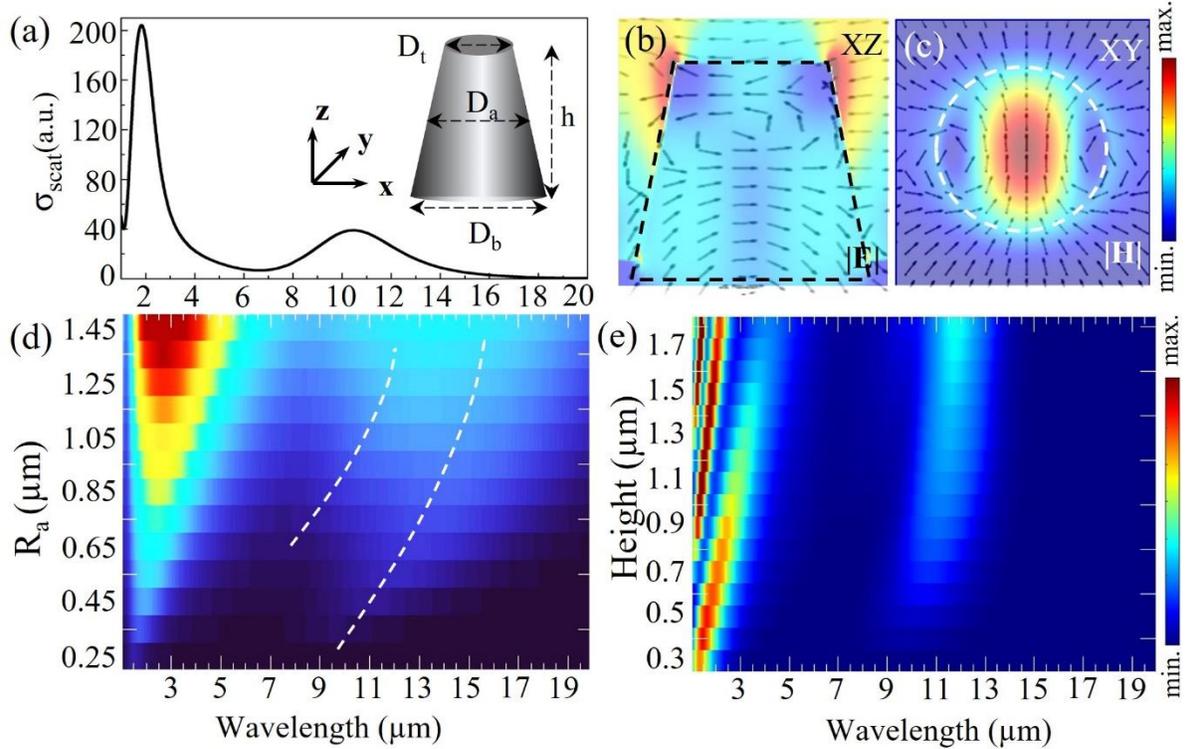

**Figure 2:** Properties of single truncated-cone $Bi_2Te_3$ resonators of $R_a = 0.44\mu m$ and $h = 0.6\mu m$. (a) The scattering spectra of a single $Bi_2Te_3$ resonator. inset shows the shape and dimensions or the resonator. (b, c) present the electric and magnetic field profiles along with their field vectors at the MD resonance ($\lambda=10.8\mu m$). (d, e) size dependent spectra and modal evolution maps as a function of the (d) average radius $R_a$ for a constant height $h=0.6\mu m$ and (e) height, h, for a constant average radius $R_a= 0.44\mu m$. White dashed lines in (d) are a guide to the eye for tracking the peaks in intensity.

To demonstrate the potential of TI $Bi_2Te_3$ material, we study the resonant response from truncated cone metasurface arrays. We start by analyzing single truncated cone resonators, followed by investigations of full metasurface arrays.

Figure 2a presents the scattering cross-section of a single $Bi_2Te_3$ resonator sitting atop of a $Bi_2Te_3$ substrate, calculated using 3D finite difference time domain (FDTD)[74]. The basic unit-cell cross-section along with the relevant dimensions are depicted in the inset of Figure 2a. In this simulation we considered the average resonator radius $R_a$, defined as $R_a = (R_b + R_T)/2$, where $R_b$, $R_T$ are the bottom and top radii of the resonator, to be $R_a = 0.44\mu m$, and the height is $h=0.6\mu m$. The scattering cross-section of the submicron resonator in Figure 2a, exhibits two pronounced resonant peaks at $\lambda=10.8\mu m$ and $\lambda=1.83\mu m$, corresponding to the fundamental and 2nd higher-order resonance modes in the MIR spectral range, respectively. Figure 2(b) and 2(c) show the electric and the magnetic field distributions in the x-z and x-y planes, respectively, at $\lambda=10.8\mu m$, and reveal the nature of the fundamental resonance. The circulating field vectors in Figure 2b are a signature of a typical magnetic dipole (MD) resonance present at λ=10.8µm wavelength. The second pronounced peak in the scattering response at λ=1.83µm is a higher-order mode mostly dominated by the electric field of the topological surface. More details on field distribution of higher order modes are given in the supporting information. The ability to support a fundamental MD resonance at λ=10.8µm wavelength in submicron resonators, demonstrates the deep subwavelength capabilities of this structure and is a consequence of the ultra-high refractive index of $Bi_2Te_3$.

Figure 2d and 2e present geometric size dispersion and modal evolution maps as a function of the average radius $R_a$ and the height, h. In the first case (Fig. 2d), when the average radius $R_a$ is increased (while fixing the height at h=0.6 µm), all resonance modes are red-shifted. For larger $R_a$ values (i.e., $R_a$ >0.65µm) the structure supports another resonant mode, manifested by splitting of the single broad MD peak into two separate peaks (both modes are marked by white dashed line in Fig. 2d, see Figure S4 in supporting information). In contrast, by increasing the height h (for fixed $R_a$ = 0.44µm), resonance peaks redshift without significant broadening. For larger height values, new higher order modes appear at lower wavelengths (for h>0.75µm). Altogether the size dispersion presented in Figures 2d and 2e, demonstrates the capability to engineer resonance wavelengths across the infrared with deep subwavelength unit cells (size <λ/10).

## Size-tunable topological metasurfaces across the MIR

As demonstrated in previous sections, high-index dielectric response of $Bi_2Te_3$ plays a significant role in generating strongly confined Mie resonance modes in deep-subwavelength unit cells. Next, we demonstrate the ability to engineer $Bi_2Te_3$ TI metasurfaces, covering the entire MIR spectral and. To understand the interplay between unit-cell size and array characteristics, and their influence on Mie resonances and modal evaluation, metasurface arrays of varying sizes and periodicities were studied. Figure 3a presents experimental and FDTD spectra of $Bi_2Te_3$ metasurfaces of average radius $R_a$= 0.44 µm and height h=0.6µm. Metasurface structures were defined using focused ion beam (FIB) milling of single crystal $Bi_2Te_3$ double-sided polished substrates. Figure 3b presents high-resolution scanning electron microscopy (SEM) image of the metasurface. Experimentally measured spectra (Figure 3a) are in excellent agreement with FDTD predictions, serving as another independent validation for the extracted optical constants (Figure 1b). The fundamental MD resonance experimentally observed at ~10.8 µm demonstrates deep subwavelength light confinement with unit cell size smaller than λ/10. This demonstration, to the best of our knowledge, is the smallest normalized unit cell (D/λ) ever demonstrated with dielectric metasurface resonators. Figure 3d presents experimental and FDTD spectra of a metasurface with a larger unit cell with $R_a$=0.975µm and h=1.178 µm (SEM in Figure 3c). All resonances are red shifted and pushed deeper into the long wavelength regime, with the fundamental MD mode emerging at λ≈16µm, demonstrating the capability to stretch $Bi_2Te_3$ metasurface response deep into the long wave IR range (limited up to λ ~18µm by our MCT detector). Figure 3e presents experimental geometric size dispersion of arrays with a constant average radius $R_a$=400nm and varying heights from h=160nm to h=720nm. Higher order mode at short wavelengths are observed across all the different spectra, continuously red-shifting from NIR wavelengths λ=1.25μm to λ=3.4 μm, whereas the fundamental MD mode is only observable when a certain height is reached, i.e., for heights larger than h≈350nm. To understand the underlining mechanism responsible for this size dispersion, we need to revisit the refractive index wavelength dispersion. The fundamental mode is a dielectric MD resonance and hence for small height values the imaginary part is still dominant, leading to significant damping of this resonance. For large enough height values, (i.e., for h>350nm in this example) the MD mode is red-shifted and pushed into the high refractive and low-loss spectral regime (λ>6μm) and will continue to increase its peak intensity as size increases. Figure 4f shows height dependent resonance wavelength shifts of experimentally extracted (dots) and FDTD calculated (solid lines) fundamental MD mode and second higher-order resonance mode. Experimentally extracted values match very well with FDTD predictions, both following the expected linear dependance and allowing to engineer resonance wavelengths with size.

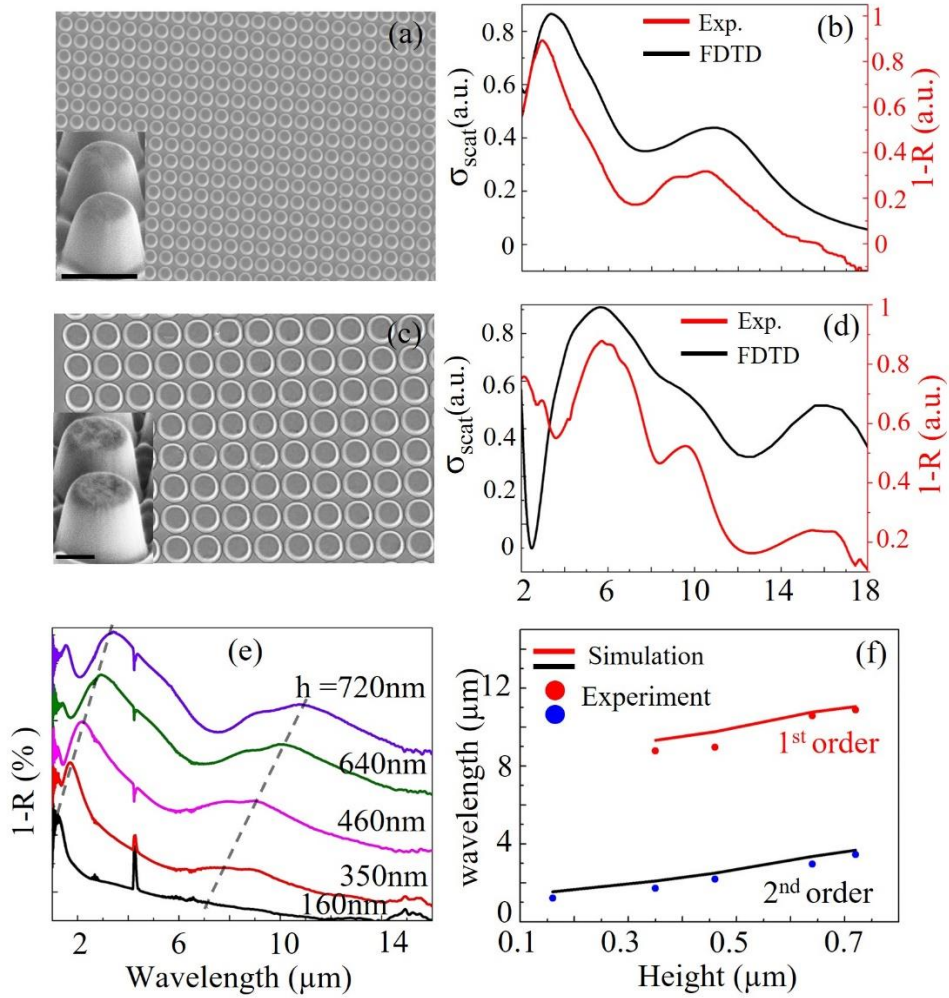

**Figure 3:** Size-tunable $Bi_2Te_3$ metasurfaces in the MIR. (a, c) present experimental (red, right y-axis) and FDTD spectra (black, left y-axis) of metasurfaces with unit cell sizes $R_a$ = 440nm h=600nm, and $R_a$ =974nm h=1.24), respectively. (b, d) corresponding SEM images of the measured metasurfaces in (a) and (c), respectively. Inset shows zoom-in view of single truncated cone unit cell. The scale bar is 1µm. (e) Experimental infrared reflection spectra of metasurfaces with variable unit cell heights and a constant average radius $R_a$=430nm ± 30nm. Spectra have been offset along the y-axis for visibility. (f) Extracted height-dependent resonance wavelength shifts of the fundamental MD (1st order) and 2nd order modes, respectively.

In previous sections we showed that $Bi_2Te_3$ metasurfaces can take advantage of the extreme refractive index of the material, providing strong dielectric response (MD resonance) in deep subwavelength meta-atoms. Next, we show how to utilize the high mobility surface states of TIs to generate large electric field enhancements. Engineering dense metasurface arrays ultimately results in enhancements of both electric and magnetic fields, simultaneously.

Figure 4a shows the metasurface design configuration and geometrical parameters used to study the effect of periodicity and inter-meta-atom interactions. Here the gap (g) is defined as $g = P-2R_a$, where P is the array periodicity. To demonstrate the effect of the gap, Figure 4b compares FDTD numerical calculations of a dense metasurface array (unit-cell dimensions $R_a$=440nm h=640nm) with g=100nm (Figure 4b, black line) and a wider-gap array with identical unit cell size, only with g=820nm (Figure 4b, red line), respectively. The spectral response of the dense array (g=100nm) includes two pronounced peaks (λ=2.83μm, and λ=10.76μm) whereas the wider gap array with g=820nm exhibits three distinct peaks (λ=1.77μm, λ=5.25μm and λ=10.76μm). The large difference in the two gaps, has negligible effect on the spectral position of the MD mode in the two metasurfaces (Figure 4b). However, when the meta-atoms are separated by a small gap (g=100nm), the two lower wavelength resonances observed (i.e., λ=1.77μm and λ=5.25μm) for moderate gap size (g=850nm), overlap and merge to form one broad peak. Figure. 4c and 4d compares the electric field distribution between the two metasurface arrays with a moderate gap g=850nm (Figure 4c) and small gap size g=100nm (Figure 4d), respectively, at the MD resonance wavelength λ=10.8μm. Clearly, smaller gap size enhances the coupling of the surface electric field between adjacent resonators, while hardly altering the electric field confined inside the resonator. This larger electric field concentration is observed in the small gap region (g=100nm, Figure 4d) with the electric field enhancement mostly originating from the high mobility surface and is 12 times larger than in the moderate gap size (g=820nm) array (see supporting information Figure S6). In should be noted, that since the gap size has negligible effect on the fields inside the resonator, the typical MD resonance field distribution inside the resonator is maintained. This effect is giving rise to magnetic field enhancement inside the resonator, due to the MD resonance, and electric field enhancement in the gap, simultaneously. More information on the effect of gap size on spectral and modal evolution is given in the supporting information. To further reveal the effect of gap on the spectral and modal characteristics, we systematically evaluated metasurface spectral and modal response as a function of the gap, as shown in Figure 4e. Continuously increasing the inter meta-atom gap, results in the splitting of the broad peak at shorter wavelength ($\lambda \approx 3\mu m$) into two distinct peaks, as also observed in Figures 4b. The spectral response of the MD resonance, however, is independent of the gap size, where only the electric field on the surface and outside the resonator vary, as demonstrated in Figures 4c and 4d. In addition, Figure 4f presents metasurface spectral response as a function of the height, for a constant inter-resonator gap (g=180nm). The spectral evolution map is similar to the single resonator response shown in Figure 2e. However, the scattering intensity in Figure 4f is significantly higher due to much stronger light-matter interactions in the dense metasurfaces.

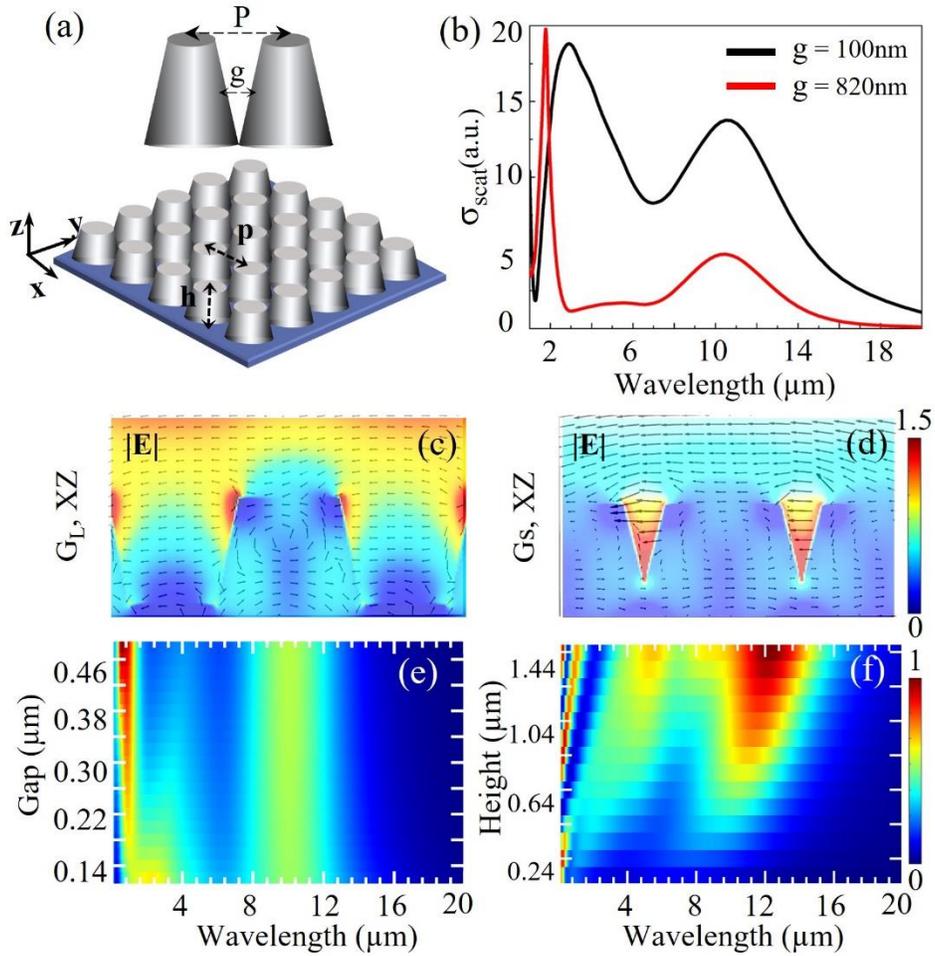

**Figure 4:** Properties of dense metasurface arrays (a) Schematic of the metasurface design parameters. (b) The scattering spectra of dense (small gap, g = 100nm) and moderate gap (g = 820nm) metasurface, respectively. (c, d) Electric field magnitude and field vectors at the MD resonance (λ=10.76μm) for small gap (g = 100nm) and moderate gap size (g = 820nm), respectively. (e, f) Spectral and modal evolution maps with respect to the gap variation at a constant height (e) and varying height at a constant gap (g=180nm) (f).

## Discussion

In this work, we extracted the optical constants of $Bi_2Te_3$ single crystals and identified a broad spectral range of high refractive index and low losses between 6 to 18µm, with peak refractive index value n≈11 at $\lambda \approx 6.3\mu m$. We demonstrate that the ultrahigh refractive index of the TI material results in deep subwavelength resonant structures with the typical largest unit cell dimension smaller than $\lambda/10$. We further demonstrate that as the inter-particle gap decreases, large electric field enhancements arise in the gap, due to the contribution of the topological surface charge. Hence, the proposed truncated-cone unit cell has significant effect in the enhancement of electromagnetic fields in the inter-particle gap, leading to metasurfaces with strong light localization that are capable of simultaneously providing large magnetic and electric field enhancements. These ultra-high index TIs may have tremendous contribution to various fields in nanophononics including spectroscopy[75,76], MIR detection[77] and manipulation of emission from quantum systems[78,79]. Altogether these results demonstrate the unique potential of chalcogenide TI as enabling material system for nanophotonics and may pave the way for applications that benefit from strong light localization and field enhancement such as ultrathin meta-optic components, non-linear generation and enhanced sensing architectures, or even more exotic applications relying on direct coupling to spin-polarized topological charges.

## Methods

**Optical characterization:**

Optical reflection measurements and extraction of optical constants were conducted on $Bi_2Te_3$ (0001), single crystals with dimensions 10 x 10 x 1.0 mm, both sides polished (MTI corporation). To derive the optical constants of single crystal $Bi_2Te_3$, we recorded broad infrared spectral measurements (1-18µm). In our experiments, we used a setup consisting of a FTIR spectrometer (Nicolet, iS50R) coupled to an infrared microscope (Nicolet, Continuum Infrared Microscope). equipped with a 20x Cassegrain objective with

NA=0.5. The permittivity and the refractive index were extracted using RefFIT software fitting of measured reflectance spectra. The analysis of the fitting to the reflection data was performed by varying the parameters of multiple Lorentz-Drude oscillators.

**FDTD simulation:** We used a commercial 3D FDTD software, Lumerical's Solutions [35] to calculate the total scattering spectra of TIs resonators and metasurfaces. In order to calculate the high-resolution electric and magnetic field distribution, we used a non-uniform three-dimensional square-shaped mesh with a minimum size of 0.2nm and a uniform overriding meshing system with the size of 5nm. This configuration of meshing was used to get higher electric and magnetic field resolution to optimize the size of the simulation. To model the resonator and substrate, we have used the experimentally extracted optical constants presented in Figure 1b. In the simulation, the Total-Field Scattered-Field (TFSF) source was used to illuminate the resonators which encapsulate the scatterer. The observation of the scattering spectrum and the magnitude of the electric and magnetic fields was calculated using a two-dimensional frequency domain field and power monitors that are placed at various locations.

**Sample Fabrication:** Metasurfaces were fabricated using dual-beam focused ion beam (FIB) milling system (Helios 5 UC, Thermo-Fisher). Structures were defined by direct milling of $Bi_2Te_3$ on the double-side polished substrates with a Ga beam with accelerating voltage of 30 kV and varying currents between 24pA for the small features (unit cell sizes of ~ 200-400nm) and up to a few nA for the largest features (unit cell sizes of ~ 3-6 µm). Typical metasurface lateral sizes were 100µm x 100µm.

## Acknowledgement

We thank the Israel Science Foundation for funding this work under grant No. 2110/19

## Conflict of Interest:

Authors declare no conflict of interest

# Supporting Information

## Deep subwavelength resonant metaphotonics enabled by high-index topological insulator $Bi_2Te_3$

*Danveer Singh[1,2], Sukanta Nandi[1,2], Yafit Fleger[2], Shany Z. Cohen[1,2] and Tomer Lewi[1,2*]*

[1]Faculty of Engineering, Bar-Ilan University, Ramat-Gan 5290002, Israel
[2]Institute for Nanotechnology and Advanced Materials, Bar-Ilan University Ramat-Gan 5290002, Israel
[*]Corresponding authors: tomer.lewi@biu.ac.il

## 1. Extraction of optical constants

### 1.1 Calculation of optical constants using refFIT

The permittivity of $Bi_2Te_3$ and the complex refractive index presented in Figure 1b were calculated using a multiple oscillator model. The model included three Lorentz and one Drude oscillators. The parameters of these oscillators were optimized to fit the experimental reflectance (see Figure 1a in the main text). The total permittivity of the system can be expressed by the sum of oscillators, as shown in Eq. 1:

$$\varepsilon = \varepsilon_\infty + \frac{\omega_p^2}{\omega^2 + i\omega\gamma_D}f_1 + f_2\left[\varepsilon_{\infty L} - 1 + \frac{\omega_{PL1}^2}{(\omega_{OL1}^2 - \omega^2) - i\omega\gamma_{L1}} + \frac{\omega_{PL2}^2}{(\omega_{OL2}^2 - \omega^2) - i\omega\gamma_{L2}} + \frac{\omega_{PL3}^2}{(\omega_{OL3}^2 - \omega^2) - i\omega\gamma_{L3}}\right] \quad (1)$$

The above equation can be compared with the generalized form, as the following:

$$\varepsilon_1 + i\varepsilon_2 = \varepsilon_{\infty,t} + \sum_{j=1}^{n} \frac{f_j}{E_j^2 - E^2 - iE\gamma_j} \quad (2)$$

Here, n is the total number of oscillators, $f_j$, $E_j$, and $\gamma_j$ are the oscillator strength, resonant energy, and linewidth of the $j^{th}$ oscillator. In our case, $j = 1$ is related to the single Drude oscillator (where $E_1 = 0$), and $j = 2, 3, 4$ are the remaining 3 Lorentz oscillators. In our fitting, we only needed two oscillator strengths $f_1$ and $f_2$ (as seen in Eq. 1) where $f_1$ is the Drude oscillator strength and $f_2$ is a fixed oscillator strength of the remaining 3 Lorentz oscillators (i.e., $f_{j=2,3,4}=f_2$). The total permittivity in Eq. 2 is $\varepsilon_{\infty,t} = \varepsilon_\infty + (\varepsilon_{\infty L} - 1)f_2$. The values of the fitting parameters are summarized in table 1.

## Table-1: Fitting parameters for the model

| Oscillators | Formula | Fitting parameters | Values |
|---|---|---|---|
|  |  | $f_1$ | 0.013 |
|  |  | $f_2$ | 0.987 |
| **Drude** | $\varepsilon_\infty + \dfrac{\omega_{pD}^2}{\omega^2 + i\gamma_D}$ | $\omega_{pD}$ (Drude part, eV) | 5 |
|  |  | $\gamma_D$ (Drude part, eV) | 0.0087 |
|  |  | $\varepsilon_\infty$ | 42 |
| **Lorentz-1** | $\dfrac{\omega_{PL1}^2}{(\omega_{OL1}^2 - \omega^2) - i\omega\gamma_{L1}}$ | $\omega_{pL}$ (eV) | 2.45 |
|  |  | $\omega_{oL}$ (eV) | 0.348 |
|  |  | $\gamma_{pL}$ (eV) | 0.375 |
| **Lorentz-2** | $\dfrac{\omega_{PL2}^2}{(\omega_{OL2}^2 - \omega^2) - i\omega\gamma_{L2}}$ | $\omega_{pL}$ (eV) | 4.6 |
|  |  | $\omega_{oL}$ (eV) | 0.789 |
|  |  | $\gamma_{pL}$ (eV) | 0.494 |
| **Lorentz-3** | $\dfrac{\omega_{PL3}^2}{(\omega_{OL3}^2 - \omega^2) - i\omega\gamma_{L3}}$ | $\omega_{pL}$ (eV) | 2.15 |
|  |  | $\omega_{oL}$ (eV) | 0.881 |
|  |  | $\gamma_{pL}$ (eV) | 0.203 |
|  |  | $\varepsilon_{\infty L}$ | 2 |

### 1.2 Extraction of optical constants using a two-layer system

As described in section 1.1, to extract the optical constants presented in Figure 1b of main text, we used a combination of multiple oscillators together with refFIT software in order to search and optimize the parameters of the oscillators to fit the experimental reflection and obtain the dielectric function. Here we use a similar physical model, where the model for the TI $Bi_2Te_3$ material is considered as a two-layer system[1]. The system consists an ultra-thin (1-2nm) Drude metal layer sits atop a (thick) bulk insulator. Such a system mimics the topological insulator where the topologically protected surface layer acts like a Drude-metal and the bulk as an insulator. The dielectric function of the insulating bulk layer is modelled by multiple Lorentz oscillators, while the thin metal layer is modelled using a Drude oscillator.

The optical constants of the two-layer system can be analytically calculated using the following equation[1]:

$$\varepsilon_{TI} = \varepsilon_\infty + i\sigma_{TI}/\varepsilon_o\omega \qquad (1)$$

here the value of the high-frequency dielectric constant is $\varepsilon_\infty = 42$, serving as a background dielectric constant that accounts for all oscillators outside of our measurement window. $\varepsilon_o$ is the vacuum permittivity and $\sigma_{TI}$ is the effective conductivity of a two-layer system. The thickness of the Drude metal is assumed to be 1.5nm and its conductivity is described by the Drude model, whereas the bulk part acts as an insulator of thickness 112nm with the conductivity described by the Lorentz model. The total conductivity of the two-layer system is described as follows:

$$\sigma_{TI} = \sigma_D f_1 + \sigma_L f_2 \qquad (2)$$

Where, $f_1 = t_{surf}/t_{total}$ and $f_2 = t_{bulk}/t_{total}$, where $t_{total}$ is the total thickness of the sample.

The conductivity of the Drude part and Lorentz part is expressed as follows:

$$\sigma_D = \varepsilon_o \frac{\omega_p^2}{\gamma_D - i\omega} \qquad (3)$$

and

$$\sigma_L = i\,\varepsilon_o\,\omega(\varepsilon_\infty - 1) + \frac{\varepsilon_o \omega_{pL}^2 \omega}{i(\varepsilon_o \omega_{oL}^2 - \omega 2) - \omega \gamma_L} \qquad (4)$$

Where, $\omega_p$ and $\gamma_D$ are the plasma frequency and damping rate of the Drude metal layer, and $\omega_L$ and $\omega_{oL}$ and $\gamma_L$ are the plasma frequency, oscillator frequency and scattering rate of Lorentz part of a two-layer system, respectively.

## 2. Permittivity and refractive index of Bi$_2$Te$_3$ single crystal

Using Equations 1 or 2 and the model parameters in table 1, the permittivity is plotted in Figure S1.

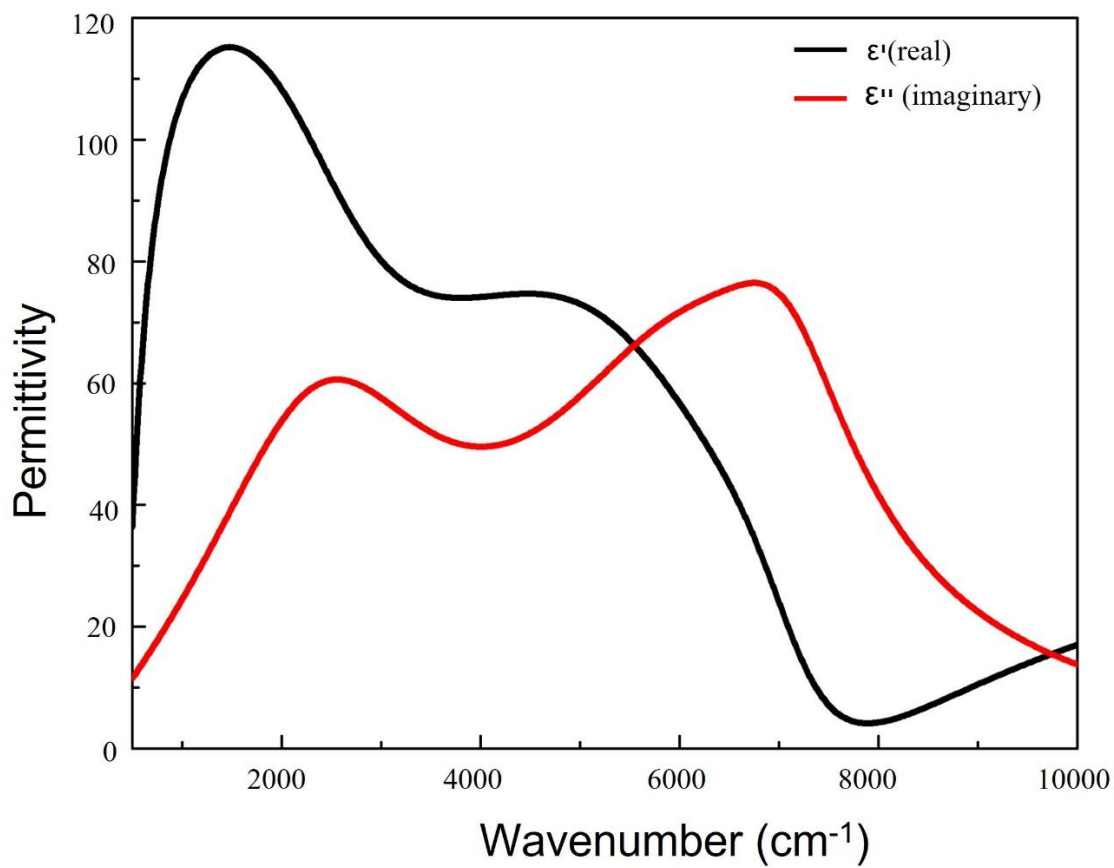

Figure S1: Dielectric function of Bi$_2$Te$_3$ crystal retrieved from fitting of the experimentally recorded reflection spectra using RefFIT software and the model parameters as summarized in table 1.

The decomposition of the total dielectric function into one Drude term – corresponding to the surface metallic state of the TI, and 3 Lorentz Oscillators – corresponding to the bulk insulator, allows us to examine their contribution. Figure S2 presents the total refractive index (blue line), and the refractive index originating from the bulk insulator only (this is done by removing the Drude term in Eq 1 or 2). From this analysis, the contribution of surface states to the Drude like decent in the refractive index, can be clearly seen. With no metallic surface present, the refractive index remains ~ constant at peak values (n≈11) for wavelengths longer than λ≈6μm. It is important to note, that his Drude term also accounts for any other free carriers originating from unintentional doping and defects in the bulk crystal.

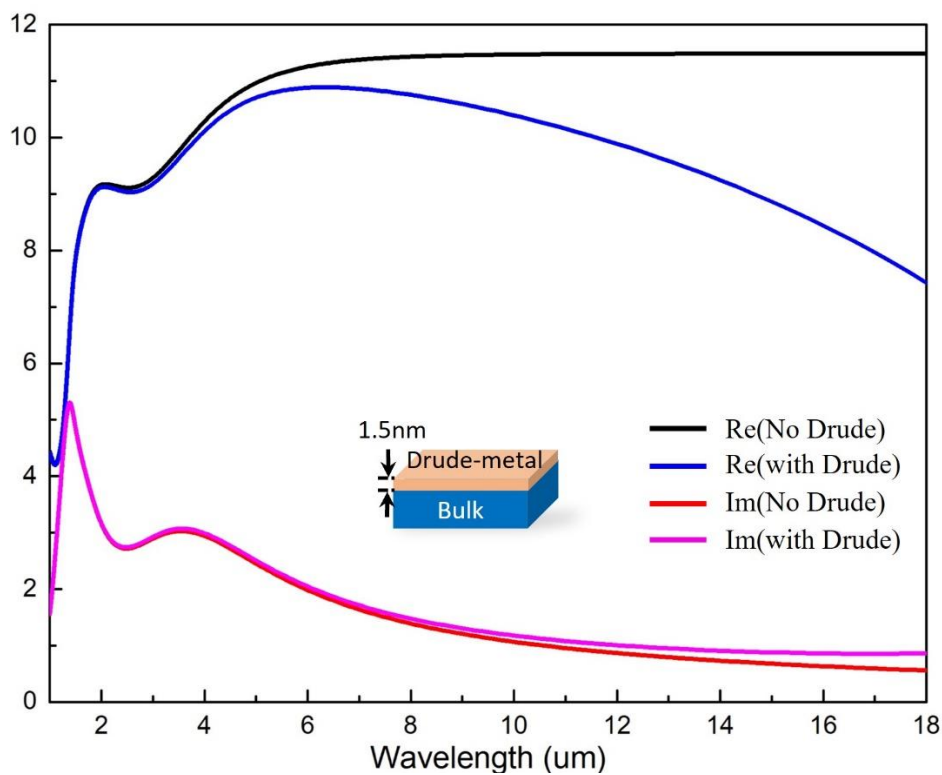

Figure S2: The complex refractive index as derived from fitting of the experimentally recorded reflection spectra using RefFIT software and the model parameters as summarized in table 1 (blue=real part, pink=imaginary part). The black and red line correspond to the real and imaginary parts of the index when the Drude contribution in removed, showing the contribution of the surface metallic state to the total response.

## 3. Spectra and higher order modes in single resonators

### 3.1 Single resonator spectra of variable average radius

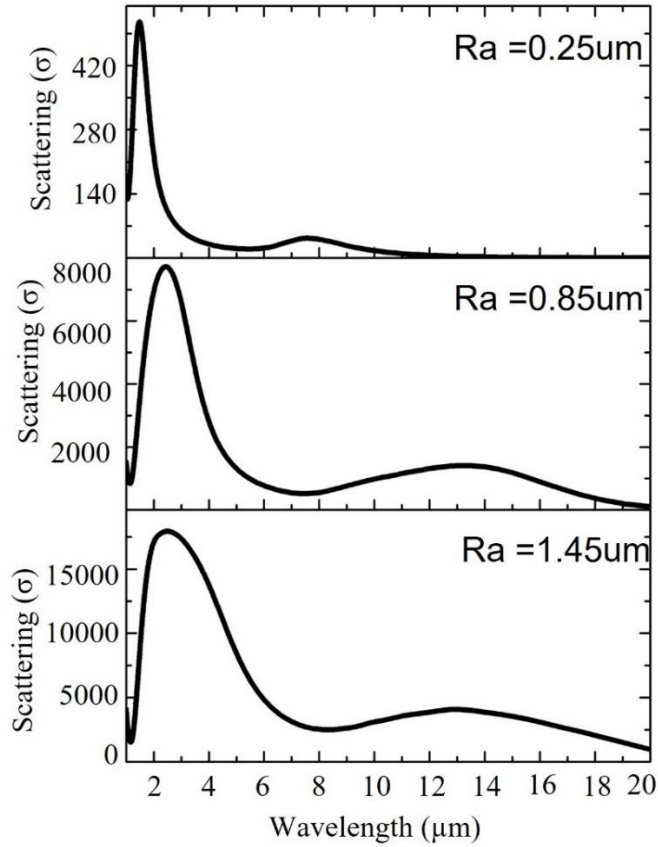

Figure S3: Scattering spectra at three different average radius values $R_a$ for single resonator at constant height h=600nm.

Figure S3 presents FDTD spectra of a single $Bi_2Te_3$ resonator placed on $Bi_2Te_3$ substrate, for 3 different $R_a$ values. These are horizontal linecuts at $R_a$=250nm, 80 and 1450nm of figure 2d of main text, respectively. Increasing the average radius $R_a$ for fixed height (h=600nm) results in red shifting of the fundamental MD mode and higher order modes. For larger $R_a$ values (i.e., $R_a$ >0.65µm) the structure supports another resonant mode, manifested by splitting of the single broad MD peak into two separate peaks, as seen in Figure 2d of main text. In figure S3 this is manifested by broadening of the long wavelength peak a result of the overlapping of the two modes (for large $R_a$ values)

## 3.2 higher order modes in single resonators

Figure S4 shows the electric and magnetic field distributions of the higher mode of the single resonator presented in Figure 2 of main text, at λ=1.83μm. The fields are mostly concentrated outside the resonator at the interface of the resonator and surrounding air (top and side walls). This is due to the optical constants at these wavelengths where $\varepsilon_1 \approx \varepsilon_2$ and k~n.

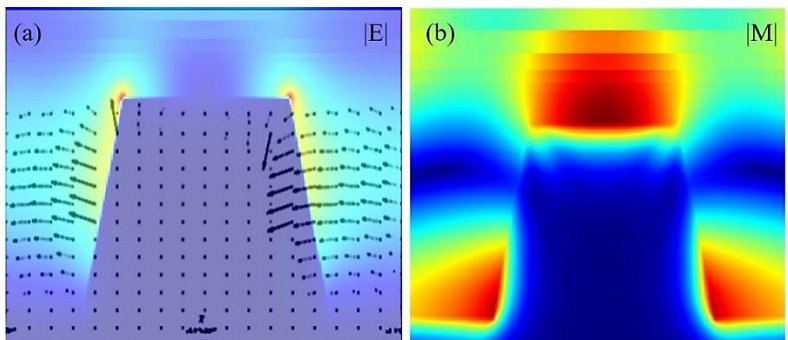

Figure S4: (a, b) Electric and magnetic magnitude of the higher order mode at λ=1.83um.

## 4. Electric field distribution of metasurface of different gap size

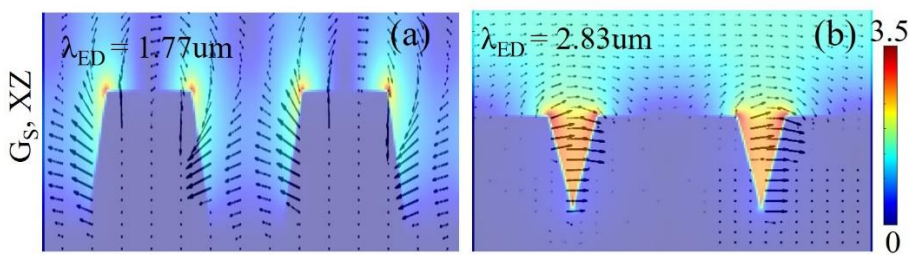

Figure S5: Electric field distribution at higher order modes in two metasurfaces configurations (a) moderate gap size g=820nm and (b) small gap size g=100nm

Figure S5 shows the electric field distributions at the higher order modes supported by metasurfaces with two different gap sizes, as discussed in the main text (Figure 4). The observed field enhancement in the gap is significantly higher in small gap sizes, and hence plays a vital role in the scattering process. The higher order mode resonance wavelength is different in both cases. In the case of a wider gap metasurface (g = 820nm), it is very close to single resonators (of the same dimensions), whereas it is red-shifted in the case of small-gap (g = 100nm), a result of two overlapping resonances as discussed in the main text (Figure 4). In the small gap case (g=100nm), large electric field enhancement resides in the gap.

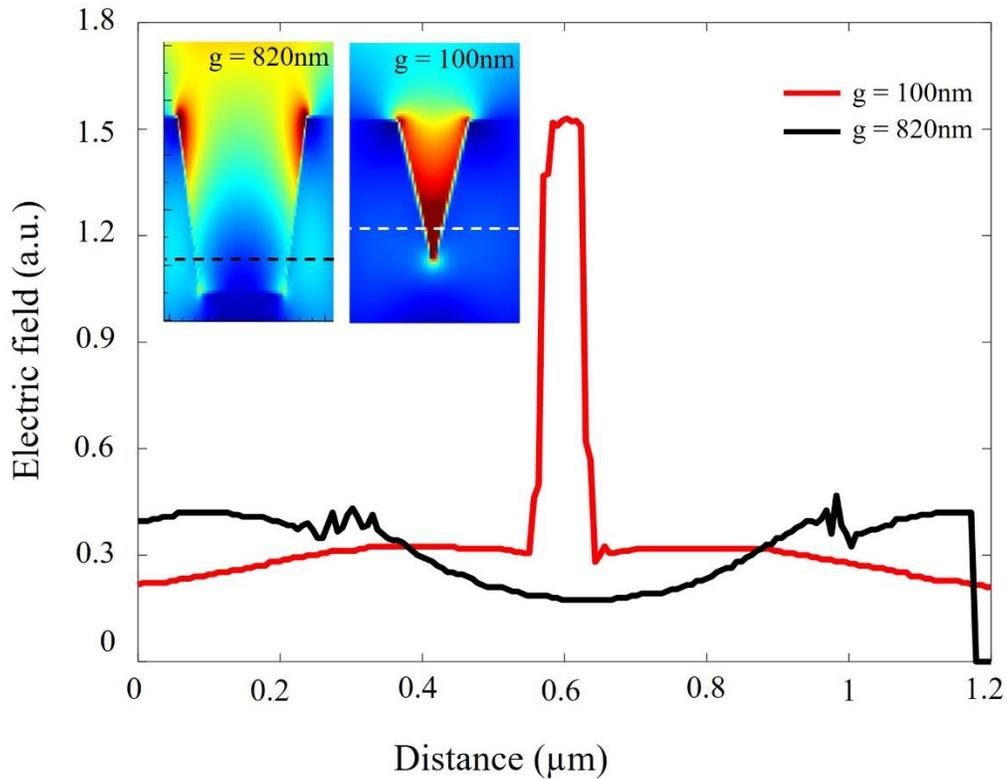

Figure S6: Electric field amplitude in two different gaps i.e., the small gap (g=100nm) and moderate gap (g=820nm), respectively. The black and red lines are the electric field amplitude profile along the linecut shown in the inset (dashed black and white lines).

Figure 6 compares electric field amplitude in the gap of the two different configurations of metasurfaces discussed in Figure 4 of main text, i.e., with unit cell size $R_a$=440nm and h=640nm, and gap sizes of g=100nm and=820nm, respectively. Larger field localization is observed in the small gap metasurface with the electric field amplitude is more than an order of magnitude larger compared with the moderate gap size g=820nm.

and Surface State of Topological Insulator Bi 2 Te 2 Se. *ACS Photonics* **6**, 2492–2498 (2019).